\documentclass[useAMS,usegraphicx,usenatbib,article,twocolumn]{mn2e}
\usepackage{multicol}
\def\apj{ApJ}
\def\mnras{MNRAS}

\def\aap{A\&A}                   
\def\apjs{ApJS}                  
\def\apjl{ApJ}                   

\def\ssr{Space Sci. Rev.}
\def\aapr{Astron. Astroph. Reviews}
\def\physrep{Phys. Reports}
\def\lsim{\;\raise0.3ex\hbox{$<$\kern-0.75em\raise-1.1ex\hbox{$\sim$}}\;}
\def\gsim{\;\raise0.3ex\hbox{$>$\kern-0.75em\raise-1.1ex\hbox{$\sim$}}\;}

\def\cmc{\rm ~cm^{-3}}

\def\kms{\rm ~km~s^{-1}}

\def\diff{\rm ~cm^2~s^{-1}}
\def \kms {\rm ~km~s^{-1}}

\usepackage{color}

\title{Non-linear model of particle acceleration at colliding shock flows}
\author[A.M.Bykov, P.E.Gladilin and S.M.Osipov]{A.M.Bykov$^{1,2}$\thanks{E-mail:
byk@astro.ioffe.ru}, P.E.Gladilin$^{1,2}$\thanks{E-mail: peter.gladilin@gmail.com} and S.M.Osipov$^{1,2}$\thanks{E-mail: osm2004@mail.ru}\\
$^{1}$Ioffe Physical-Technical Institute of the Russian Academy of Sciences,Saint-Petersburg, Russia,\\
$^{2}$Saint-Petersburg State Polytechnical University,Saint-Petersburg, Russia}

\begin{document}

\date{Accepted 2012 December 4.~~~ Received 2012 September 18}


\maketitle

\label{firstpage}

\begin{abstract}
Powerful stellar winds and supernova explosions with intense energy
release in the form of strong shock waves can convert a sizeable
part of the kinetic energy release into energetic particles. The
starforming regions are argued as a favorable site of energetic
particle acceleration and could be efficient sources of nonthermal
emission. We present here a non-linear time-dependent model of
particle acceleration in the vicinity of two closely approaching
fast magnetohydrodynamic (MHD) shocks. Such MHD flows are expected
to occur in rich young stellar cluster where a supernova is exploding
in the vicinity of a strong stellar wind of a nearby massive star. We find
that the spectrum of the high energy particles accelerated at the
stage of two closely approaching shocks can be harder than
that formed at a forward shock of an isolated supernova remnant. The
presented method can be applied to model particle acceleration in a variety of
systems with colliding MHD flows.
\end{abstract}

\section{Introduction}
Diffusive shock acceleration (DSA) mechanism thought to be operating
at the fast shocks of young supernova remnants (SNRs) is  likely
responsible for the galactic cosmic ray (CR) acceleration up to the
"knee-region" energies (see e.g. \citet[][]{Hillas05,AhaByk12}) and
 may even exceed $10^{17}$ eV as it was advocated by
\citet[][]{ptu10} for the case of nuclei accelerated in an isolated
young Type IIb SNRs. The basic features of the DSA process have been
revealed in the pioneering papers of \citet[][]{axf77},
\citet[][]{krym77}, \citet[][]{bell78} and \citet[][]{bo78} where
the high efficiency of the acceleration mechanism was shown. Effects
of nonlinear backreaction of the accelerated particles on the
structure of the supersonic shock flow were discussed later by
\citet{dv81}, \citet{be87}, \citet{berkrym88}, \citet{bell87},
\citet{je91}, \citet{MalDru01}, \citet{blasi04}, \citet{AB05},
\citet{VladByk08} and \citet{revilleea09}. It has been found that
the pressure of accelerated particles can modify the bulk plasma
flow in the shock upstream and that may result in a substantial
increase of the flow compression and flattening in the particle
spectra at the maximal energies. Magnetic field fluctuations in the
shock vicinity may be highly amplified by instabilities driven by
the cosmic ray current and CR-pressure gradient in the strong shocks
\citep[e.g.][]{bell04,boe11,schureea12}. That is an important factor
to determine the highest energies of particles accelerated by
shocks.

The maximum energy of accelerated particles strongly differs for
different types of supernovae (see e.g. \citet[][]{ptu10}), it
depends on the circumstellar medium around the supernova progenitor
star. Moreover, core-collapsed supernovae produced by massive stars
often occur in OB-star associations where the intense radiation of
hot massive stars, powerful stellar winds and supernova shocks
strongly modify the interstellar environment, producing large hot
cavities of a few tens of parsec size, called superbubbles. For
instance, the Carina OB  stars complex contains about 70 O-type
stars and more than hundred B0-B3 stars confined in a region of
about 40 pc size \citep[][]{carinaOB11}. Recently, Fermi telescope
detected an extended cocoon-shape source of gamma-ray emission
associated with a massive star-forming region Cygnus OB2
\citep[][]{fermiSB11}. The detection indicates the presence of
active particle acceleration processes in the association. Cygnus
OB2 contains over 50 O-type stars and hundreds of B-type stars
\citep[][]{cygnusOB10}. Compact sources like binary massive stars
(with the stars separated by a few astronomical units) may
accelerate relativistic particles (\citet[][]{eu93,pittard06}) on a
month time scale.  Collective emission of such binaries may
contribute to the gamma-ray flux observed by Fermi. However, if the
detected gamma-ray emission is truly extended, it can be attributed
to relativistic particles accelerated by multiple shocks in the
superbubble, as modeled by \citet[][]{b01}, \citet[][]{bt01} and
\citet[][]{fermarcow10}. Multiple successive interactions of
particles with large shocks and rarefactions of a superbubble scale
size occur on the time scale of about 10$^5$ years.

In the present paper we model a class of a few parsec size particle
accelerators associated with collision of a young supernova shock
with a fast stellar wind of a massive star.  The modelled stage
starts a few hundred years before the supernova shock collides with
the wind termination shock as it is illustrated in Figure
\ref{fig1}. At this stage the maximal energy particles accelerated
via DSA at the SNR shock  reach the fast wind termination shock and
are scattered back by magnetic fluctuations carried by the fast
stellar wind. Therefore, the high energy particles that have mean
free path $\Lambda(p)$ larger than the distance between the two
shocks $L_{12}$ start to be accelerated by the converging fast
flows. This is the most favorable circumstance for the efficient
Fermi acceleration. While the structure of the MHD flow in the
vicinity of a supernova shell colliding with the stellar wind
termination shock is rather complex, it is possible to consider a
simplified model that captures the basic features of the
acceleration process. When the magnetic field fluctuations are
strong enough to provide the so-called Bohm diffusion regime with
$\Lambda(p) = \xi r_{g}(p)$, where $p$ is the particle momentum,
$r_{g}(p)$ is the particle gyroradius in the mean magnetic field and
$\xi \gsim$ 1, the high energy particles bouncing between the
converging flows are likely to have a spectrum harder than that
produced at an individual shock and may contain a sizeable fraction
of the total kinetic energy of the converging MHD flows. Thus, we
formulate a simplified nonlinear approach to account for the effect
of the flow modification by the pressure of accelerated CRs. We
consider the system evolution stage when radii $R_{\rm sh}$ and
$R_{\rm sw}$ of the two shocks are much larger than the distance
$L_{12}$ justifying a local one-dimensional approach.

In \S~\ref{FEB} we generalize for the case of two shock flows the
semi-analytical model originally developed by \citet{blasi04},
\citet{AB05} and \citet{Cap10} to account for nonlinear CR
modification of a single stationary strong shock flow. A two shock
flow can be stationary in the case of a supersonically moving star
with a fast wind. In that case the bow shock and the termination
shock are steady in the rest frame of the moving star. On the
contrary in the case of a supernova shock approaching the stellar
wind termination shock the system is non-steady. Therefore to model
the system we constructed in \S~\ref{time} a simplified time
dependent approach valid for the shocks with fast CR acceleration.
If magnetic field amplification provides the efficient Bohm
diffusion resulting in fast CR acceleration, one can implement
CR-pressure-modified profile of the instant
 local flow obtained in the semi-analytical solution
into the time-dependent CR transport equation.

\section[]{Semi-analytic nonlinear model with free escape boundary}\label{FEB}
Consider a model describing a population of high energy CR
particles with $\Lambda(p) > L_{12}$ in a vicinity of two approaching shocks
with $R_{\rm sh}= R_{\rm sw} \gg L_{12}$. For simplicity and the
sake of brevity we will obtain here the solution for the shocks with
the same parameters (velocity profile and hydrodynamic quantities) but this model
can be easily generalized.

\begin{figure}
\resizebox{10cm}{13cm}{
\includegraphics{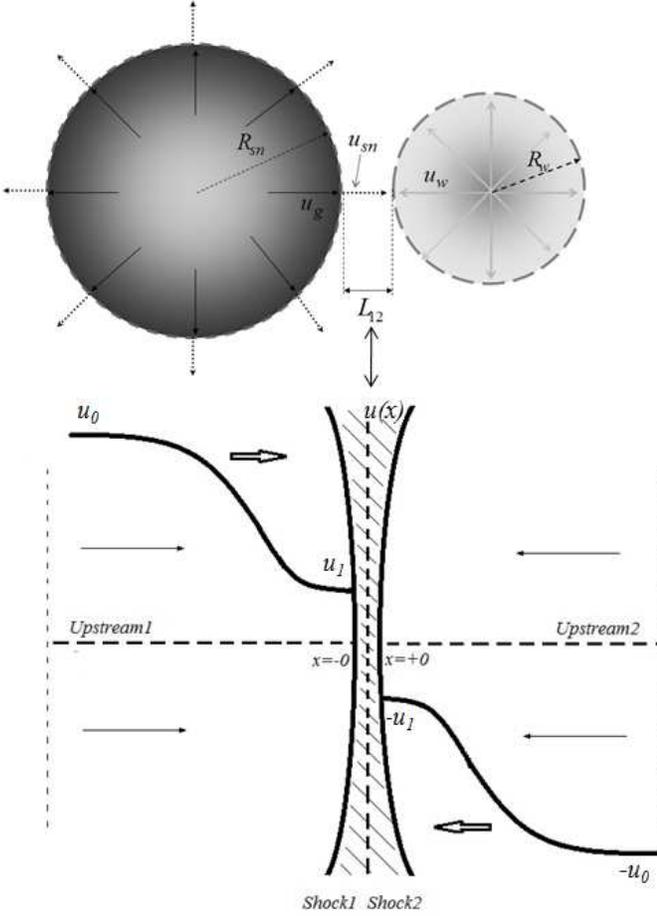}}
\caption{Simplified scheme of velocity profiles for the approaching
shocks system.} \label{fig1}
\end{figure}

In Figure~\ref{fig1} we illustrate a simplified scheme of the local
flow with two approaching shocks. At $x<0$ there is the upstream
region of the shock $1$ and at $x>0$ there is the upstream region of
the shock $2$. Shock fronts are situated at the $x=0^{-}$ and
$x=0^{+}$ positions and the CR free escape boundaries (FEBs) are at
$x=-x_0$ and $x=x_0$. The CR FEB condition requires that
\begin{equation}
f(-x_0,p)=f(x_0,p)=0,
 \label{eq:feb_condition}
\end{equation}
where $f(x,p)$ is a stationary CR distribution function.
 To derive the distribution function at the shock in
the case of the two colliding shock fronts we employ a steady-state
diffusion-convection equation with the CR particle injection rate
$Q(x,p)$
\begin{eqnarray}
u(x)  \frac{\partial f (x,p)}{\partial x} - \frac{\partial}{\partial
x} \left[ D(x,p)  \frac{\partial}{\partial x} f(x,p) \right] =
\nonumber \\ = \frac{p}{3} \,\frac{d u(x)}{d x}\,\frac{\partial
f(x,p)}{\partial p} + Q(x,p)\delta(x). \label{eq:transport}
\end{eqnarray}

where $D(x,p)$ is the diffusion coefficient,
$u(x)$ is the fluid velocity. Integrating Eq.~\ref{eq:transport}
from $x=-x_0$ to $x=x_0$ and splitting integration region into 3
parts: from $x=-x_0$ to $x=0^{-}$, from $x=0^{-}$ to $x=0^{+}$ and
from $x=0^{+}$ to $x=x_0$, one can achieve the following equation
$$
\frac{1}{3} p \frac{\partial}{\partial p} \left[\int_{-x_0}^{0^{-}} + \int_{0^{+}}^{x_0} \right] f \frac{du(x)}{dx} dx + \left[\int_{-x_0}^{0^{-}} + \int_{0^{+}}^{x_0} \right] f \frac{du(x)}{dx} dx +
$$
\begin{equation}
 +2 (   \phi_{esc}(p) -\frac{1}{3} p \frac{\partial f_{0}}{\partial p} u_{1})+Q_{0}(p) =
 0,
\end{equation}
where $f_{0}(p)=f(x=0,p)$  is the distribution function at the
shock, $Q_{0}(p)=Q(0,p)$  is the injection rate at the shock and
$\phi_{esc}(p)=-\left [D(x,p)\frac{\partial f}{\partial x}\right
]_{- x_0}$  is the escaping flux of energetic particles at $x=-x_0$
and $x=x_0$.

Assuming the symmetry of the shock flow profiles and introducing
\begin{equation}
u_p = u_1 - \frac{1}{f_0(p)}
\int_{-x_0}^0 dx (du/dx)f(x,p)\ ,
\label{eq:up}
\end{equation}
where $u_1$ is the fluid velocity
immediately upstream (at $x=0^-$ and $x=0^+$), we obtain:
\begin{equation}
p \frac{\partial f_{0}}{\partial p} = - \frac{3}{u_{p}} \left\{f_{0} \left(u_{p} + \frac{1}{3} p \frac{\partial u_{p}}{\partial p}\right) -\phi_{esc}(p) - \frac{1}{2} Q_{0}(p) \right\}.
\end{equation}
The function $u_p$ is instrumental to account for the nonlinear
modification of the flow due to backreaction of the accelerated
particles and it is included into the non-linear calculations
\citep[for details see e.g.][]{AB05}.

Thus the distribution function can be evaluated as
\begin{equation}
f_{0}(p) = \int_{p_{inj}-0}^{p} \frac{d
\overline{p}}{\overline{p}}\frac{3 (\phi_{esc}(\overline{p}) +
Q_{0}(\overline{p})/2)}{u_{\overline{p}}}\,K(p,\overline{p}),
\end{equation}
where
\begin{equation}
K(p,\overline{p})=
\exp\left[-\int_{\overline{p}}^{p}\frac{dp^{'}}{p^{'}}
\frac{3}{u_{p^{'}}} (u_{p^{'}} + \frac{1}{3} p^{'} \frac{\partial
u_{p^{'}}}{\partial p^{'}}  ) \right].
\end{equation}
The CR injection rate in Eq.~(\ref{eq:transport}) can be written as
\begin{equation}
Q(x,p) = \frac{\eta n_{gas,1} u_1}{4\pi p_{inj}^2}
\delta(p-p_{inj})\delta(x),
\end{equation}
where $p_{inj}$ is particle injection momentum, $n_{gas,1}=\rho_0 u_{0}/m_{p} u_{1}$
is the gas density immediately upstream ($x=0^-$ and $x=0^+$), $n_0$ and $\rho_0$ are
gas density and mass density at the shock and $\eta$ is the fraction
of the particles which are considered to be
injected into the acceleration process. Hence, $f_{0}(p)$ can be
presented in a simple form
\begin{equation}
f_{0}(p) = \frac{3 \eta \rho_{0} }{8 \pi m_{p} U_{p}(p)}
\frac{1}{p^{3}} - \frac{3}{U_{p}(p) p^{3}} \int_{p_{0}}^{p}
\phi_{esc}\, p^{'2} dp^{'}, \label{eq:solution}
\end{equation}
where $U_{p}\equiv u_{p}/u_{0}$.

Eq.~(\ref{eq:solution}) represents the momentum distribution
function at the high-energy limit ($\Lambda(p)>L_{12}$) for the MHD
flow with two identical colliding shocks with the FEBs. Note, that
for every momentum $p$ function $f_{0}(p)$ is proportional to
$p^{-3}$ with correction factor $U_{p}(p)$. The first term in
Eq.~(\ref{eq:solution}) reflects CR injection and the second term is
due to the escaping flux. The second term is most important for the
CR spectral shape at the highest energies (i.e. above $p>p_{\star}$
where $\Lambda(p_{\star})=L_{12}$). Such a simple form of the CRs
distribution function can be obtained in the case of the symmetric
flow. The corresponding expression for the general case is somewhat
more complex and will be discussed elsewhere.

\section{An approximate solution of diffusion-advection
equation}\label{DAEq}
In this section we use an approximate solution of the
one-dimensional diffusion-advection equation proposed by
\citet{Cap10} with the distribution function at the shock $f_0(p)$ for the
case of two converging shocks derived in the previous section. The
solution was obtained by integrating the diffusion-convection
equation from $-x_0$ to an arbitrary point $x$ in the
shock upstream, with the FEB condition as given by
Eq.~(\ref{eq:feb_condition}). A symmetric model is considered, so
that the solution in the upstream region $1$ ($x<0$) does not
differ from the solution in the upstream region $2$ (see
Fig.~\ref{fig1}). Following \citet{Cap10} we used an
approximation of the exact solution to the transport equation
\begin{equation}\label{eq:app}
f(x,p)=f_0\exp\left[-\int_{x}^{0} dx'\frac{u(x')}{D(x',p)}\right]   \left[ 1-\frac{W(x,p)}{W_{0}(p)}\right],
\end{equation}
\begin{equation}
\phi_{esc}(p)=- \frac{u_{0}f_0}{W_{0}(p)}\,, \label{eq:esc}
\end{equation}
where $D(x,p)$ is the CR diffusion coefficient,
\begin{equation}
W(x,p)=u_{0}\int_{x}^{0}  dx' \frac{\exp\left[-\psi(x',p)\right]}{D(x',p)},
\end{equation}
\begin{equation}
\psi(x,p)=-\int_{x}^{0} dx'\frac{u(x')}{D(x',p)}\ ,
\end{equation}
and $W_{0}(p)=W(x_0,p)$.

These expressions are exact in the test-particle limit, as one can
easily verify. The iterative method that has been used in the
calculations  is based on the successive approximations to the
solution $f(x,p)$ that satisfy both the CR transport and the
momentum-energy conservation equations.
 The momentum conservation equation, normalized to $\rho_{0}u_{0}^{2}$ reads
\begin{equation}\label{eq:momentum}
U(x)+P_{c}(x)+P_{w}(x)+P_{g}(x)=1+\frac{1}{\gamma M_0^2}\,,
\end{equation}
where $M_0$ is the Mach number of the unperturbed flow. The
normalized cosmic ray pressure
\begin{equation}\label{eq:pc}
    P_{c}(x)=\frac{4\pi}{3\rho_0u_0^2}\int_{p_{inj}}^{\infty}dp~ p^3~ v(p) ~f(x,p)\,,
\end{equation}
where $v(p)$ is the velocity of the particle. The normalized
pressure of magnetic fluctuations generated via the resonant
streaming instability \citep[see Eq.(42) in][]{Cap09}
\begin{equation}\label{eq:pw}
    P_{w}(x)=\frac{v_A}{4u_0}\frac{1-U^{2}(x)}{U^{3/2}(x)},
\label{eq:pw}
\end{equation}
where $v_A=B_0/\sqrt{4 \pi \rho_0}$ - is the Alfven velocity, $B_0$
- is the strength of the unperturbed magnetic field. Equation
~(\ref{eq:pw}) was obtained from the stationary equation for
 growth and transport of magnetic turbulence. We used the
approximation for the turbulent energy flux  $F_w(x)\simeq
3u(x)p_w(x)$, assuming that $v_a \ll u(x)$ \citep[see,
e.g.,][]{Cap09}.
 The normalized pressure of the background gas
\begin{equation}\label{eq:pg}
P_{g}(x)=\frac{U^{-\gamma}(x)}{\gamma M_{0}^{2}}\,,
\end{equation}
where $U(x)=u(x)/u_0$ and $\gamma$ is the adiabatic index.

The first iteration starts from a guess value for
$U_{1}=u_1/u_0$, which uniquely determines $P_{w1}$, $P_{g1}$
and $P_{c1}$ via Eqs.~(\ref{eq:pw}), (\ref{eq:pg}) and
(\ref{eq:momentum}).

To the first approximation we start with a test-particle guess for
$f(x,p)$, properly normalized in order to account for the pressure
$P_{c1}$, and calculate $P_{c}(x)$ from Eq.~(\ref{eq:pc}) and then
$U(x)$ with Eq.~(\ref{eq:momentum}). Then the updated velocity
profile is used to construct a new $P_{w}(x)$ and  $\delta
B(x)=\sqrt{8\pi\rho_{0}u_{0}^{2}P_{w}(x)}$ which is employed to
update the diffusion coefficient $D(x,p)=v\,p\,c/3e\,\delta B(x)$.

According to Eq.~(\ref{eq:app}), a new profile of $f(x,p)$ is
obtained with the initial distribution function and the new $U(x)$
and $D(x,p)$. The procedure is iterated until a convergence is
reached, i.e.\ until the remainders of $f(x,p)$ and its
normalization factor reach a preset accuracy between two steps.

For an arbitrary value of $U_{1}$ with the fixed list of the model
parameters, however, the required normalization factor can differ
from 1, then the process is restarted with another choice of $U_{1}$
until no further normalization is needed. The distribution function
calculated with the value of $U_1$ obtained by this approach is, by
construction, the solution of both diffusion-convection and
conservation equations for the two shock flow model.

\begin{figure}
\resizebox{10cm}{8cm}{
\includegraphics{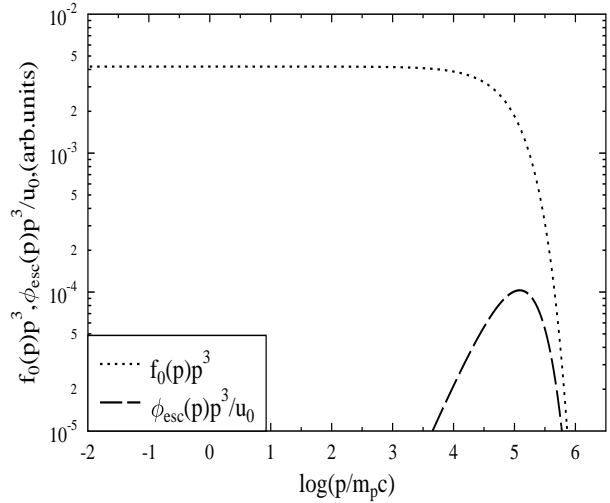}}
\caption{The CR proton spectrum $p^{3}f_0(p)$ at the shock position
(dotted line) and the escaping flux $p^{3}\phi_{esc}(p)/u_0$ (dashed
line), simulated in the non-linear semi-analytical model of DSA in
the colliding shock flow}
\label{fig:fig2}
\end{figure}

In Fig.~\ref{fig:fig2} we illustrate the CR spectrum  in the limit
of small $L_{12}$ when the shocks are colliding. The proton
distribution function at the shocks (dotted line) given by
Eqs.~(\ref{eq:solution})-(\ref{eq:app})
 and the corresponding escaping flux (dashed line) given by Eq.~(\ref{eq:esc})
were calculated with the semi-analytical approach. The following parameters
were chosen for the calculation:  the shock velocity
 $u_0=3,000~\kms$, the free escape boundary is at
$1.0$ pc on the both sides from the shocks, the background gas
density is $n_0=1.0~\cmc$ and the Alfven velocity is $v_A=30~ \kms$.
Within the spectrum calculation we used the
diffusion coefficient $D(x,p)$=5$\times$10$^{20}\,\,[p /(GeV/c)]
\times (\delta B (x) /100 \mu G)^{-1}$ $\diff$.
 The ratio of the escaping and injection
fluxes is governed by the value of the diffusion coefficient in the
upstream region $D(x,p)$, and in the model it is given by the
expression
\begin{equation}
\frac{\phi_{esc}(p)} {u_0 f_0(p)}=\frac{1}{W_0(p)}.
 \label{eq:phi_f}
\end{equation}
At the maximal momenta of the accelerated particles the dimensionless
ratio Eq.~(\ref{eq:phi_f}) scales as  $D(x_0,p)$.

Note that the proton spectrum at the shock is not strongly modified
by the cosmic ray pressure and differs from the case of a single
modified shock simulated by \citet[][]{Cap10} with the same
injection parameter $\xi_{inj}=3.3$, corresponding to a fraction of
injected particles $\eta \simeq 1.2 \times 10^{-3}$.

It is worth noting that the spectral shape of the escaping CR flux
(showed as the dashed line in Fig.~\ref{fig:fig2}) differs from that in the case
of a single strong shock. In the case of the approaching shocks flow
the escaping particles form a flattened spectra at lower energies.
The spectral shape is no longer symmetric and it departs from the
parabolic law.

In the case of a SN shock approaching a termination shock of the
stellar wind, the system is non-steady.  Nevertheless, one can make
use of the non-linear {\sl steady} solution given above to simulate
the {\sl instant} flow profiles between the shocks, provided the CR
acceleration time $\tau_{a}$ is short compared the dynamical time
scale $\tau_{\rm dyn} = L_{12}/u_{sn}$, where $u_{sn}$ is the
velocity of the SNR forward shock. The simulated instant profiles
can be then implemented into the non-steady transport equations for
CR protons and electrons to account for the time-dependent position
of the approaching shocks.

It is important to note that the plasma flow in between the
converging shocks is governed by the {\sl gradient} of the CR
pressure. The CR pressure itself may be dominated by the highest
energy part of the CR spectra. However, the CR protons of momenta
$p>p_{\star}$  propagate in between the shocks without scattering,
and their pressure is nearly homogeneous. In result, the flow is
modified by the pressure gradient of the CR particles with $p <
p_{\star}$. These have the acceleration time shorter than $\tau_{\rm
dyn}$ and are mainly concentrated in the vicinity of the shocks.

\section{Time-dependent model simulations}\label{time}
The non-linear modification of the colliding shock flow by the
accelerated cosmic ray pressure can be accounted for within a
time-dependent model. A specific feature of the simulation is that
the highest energy CRs of $p_{\rm max} \geq  p \gsim p_{\star}$
propagate without scattering, but their momenta are still nearly
isotropic being scattered in the extended shock downstream regions.
To treat the case one can use the transport equation in the form of
the telegraph equation. The telegraph equation  can be derived from
the Boltzmann kinetic equation for a nearly-isotropic CR particle
distribution \citep[see, e.g., Eq.(35) in][]{earl74}. The equation
allows a smooth transition between the diffusive and the
scatter-free propagation regimes. With time-dependent simulations we
calculated the energy spectra of cosmic ray protons and electrons at
a SNR shock approaching a strong wind of a nearby early type star.
For relativistic electrons/positrons we account for the energy
losses due to synchrotron and inverse Compton (IC) radiation.

We solve one-dimensional transport equations for the
pitch-angle-averaged phase space distribution function of protons,
$f_p(x, p, t)$, and electrons, $f_e(x, p, t)$,  given by
\begin{eqnarray}\label{difadv_p}
\tau(p)\frac{\partial^2 g_p}{\partial t^2} + \frac{\partial
g_p}{\partial t}+u(x)\frac{\partial g_p}{\partial x} -
\frac{1}{3}\frac{\partial u(x)}{\partial x} \left( \frac{\partial
g_p}{\partial y}- 4 g_p\right) =  \nonumber \\
\frac{\partial}{\partial x}\left( D(x,p) \frac{\partial
g_p}{\partial x}\right),
\end{eqnarray}
\begin{eqnarray}
\tau(p)\frac{\partial^2 g_e}{\partial t^2} + \frac{\partial
g_e}{\partial t}+u(x)\frac{\partial g_e}{\partial x}-
\frac{1}{3}\frac{\partial u(x)}{\partial x} \left( \frac{\partial
g_e}{\partial y}- 4 g_e\right) = \nonumber \\
\frac{\partial}{\partial x}\left( D(x,p) \frac{\partial
g_e}{\partial x}\right)+ \exp(y)\frac{\partial}{\partial y}\left[b
\exp(-2y)g_e\right], \label{difadv_e}
\end{eqnarray}
where $g_p = p^4f_p$, $g_e = p^4f_e$, $y = ln(p)$. The dimensionless
particle momentum $p$ is expressed in the units of $m_pc$. The
cooling term $b(p) = -dp/dt$ describes the electron synchrotron and
IC losses  \citep[see, e.g.,][]{kang11}. Here $\tau(p) =
\Lambda(p)/v$ is a CR particle pitch-angle scattering mean free
time. The transport equations Eq.~ (\ref{difadv_p}) and Eq.~
(\ref{difadv_e}) are modified telegraph equations \citep[see,
e.g.,][]{earl74,Topt85}. The telegraph equation embodies both the
diffusive propagation of low energy particles $p < p_{\star}$ with
the mean free paths below $L_{12}$ and the ballistic scatter free
propagation of particles with $p \gsim p_{\star}$. The transport
equations are valid for nearly isotropic particle distributions that
can be maintained for particles with momenta $p_{\rm max} \geq  p
\gsim p_{\star}$ that are scattered in the inflowing plasma
(Upstream 1 and 2 regions shown in Fig.~\ref{fig1}). At high
energies  $p \gsim p_{\star}$ the ballistic  propagation of CR
particles in between the shocks is governed by the first terms of
the transport equations. The diffusion coefficient in the regime
saturates at $D_{is} \approx cL_{12}$.

At the moving shocks we applied the standard matching conditions
used in DSA \citep[see, e.g.,][]{bell78,be87}. The matching
conditions are equating the isotropic parts of the distribution
functions and the particle fluxes in the phase space at the shock
surface to guarantee the particle number conservation. At the
simulation box boundaries $x = \pm x_b$ we used the free escape
boundary conditions $g_{i1}(t,-x_b,p)=0$ and $g_{i2}(t,x_b,p)=0$.
The simulation starts with the initial test particle distribution
function at the shocks $dN(p)/dp\sim 1/p^2$ and
$\partial{N}/\partial{t} = 0$ at $t=0$.

The transport equations were solved with integration-interpolation
algorithm and the standard Crank-Nicolson scheme. The model allows
to calculate $f_{e,p}(x,p,t)$ at any position between the shocks and
in the post-shock flows.

\begin{figure*}
\center
\resizebox{17cm}{22cm}{
\includegraphics{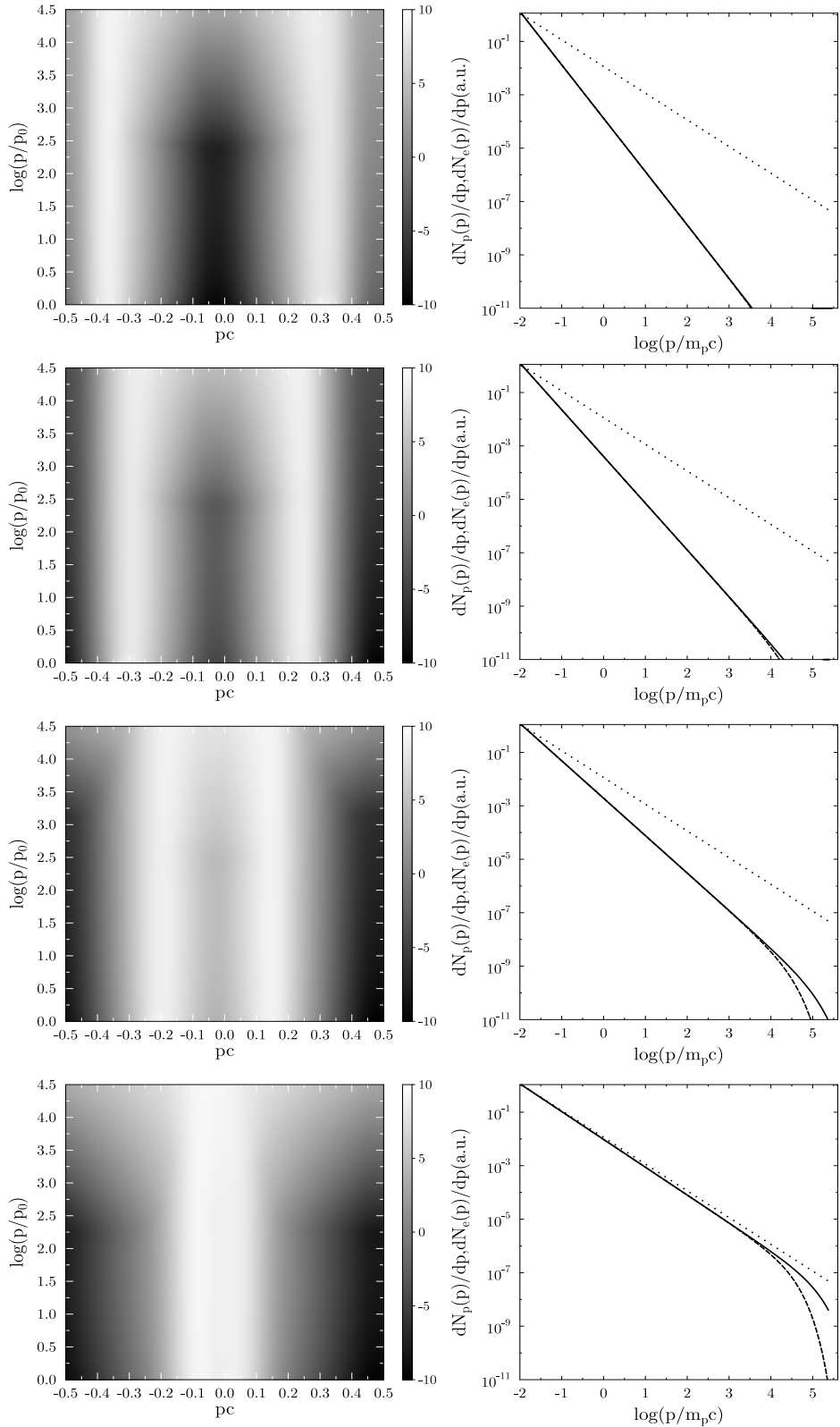}}
\caption{On the left: the proton distribution function $f(x,p)\,p^3$
(gray scaled) as a function of the CR momentum and the position $x$,
presented for four different distances between the shocks - from top
to bottom: 0.6 pc, 0.5 pc, 0,3 pc, 0.1 pc. The left shock is
propagating in the positive direction of the $x$- axis and the right
shock is moving in the opposite direction. The shock velocities are
3,000 $\kms$. On the right: the proton and electron  spectra at the
moving left shock for the times corresponding to the four intershock
distances given above. Solid line - simulated proton spectrum
$dN_p/dp$, dashed line - simulated electron spectrum $dN_e/dp$
multiplied by a factor of 100 for the clarity of presentation. The
asymptotic spectral shape
 $dN/dp\propto1/p$ is shown with dotted line.} \label{fig:fig3}
\end{figure*}

In Fig.~\ref{fig:fig3} (left panels) we present the result of
calculations of the proton distribution function. This was obtained
in the time-dependent model simulations of the flow between
colliding shocks. Four panels display $f_p(x,p,t)$ in the phase
space $(p,x)$ at four different times which correspond to the
inter-shock distances of 0.6 pc, 0.5 pc, 0.3 pc and 0.1 pc. The free
escape conditions are applied at the simulation box boundary at
$x_b$ = 0.55 pc. At each time step the flow model implements the
modified velocity profiles, u(x,t), obtained in the semi-analytical
iteration scheme described in paragraph \S3. The four right panels
in Fig 3 show the proton and electron spectra, $dN_p(p)/dp$ and
$dN_e(p)/dp$, at the moving left shock, at the corresponding time
moments, compared to the expected spectrum $dN/(p)/dp \propto 1/p$.
In Fig.~\ref{fig:fig3} the spectral evolution of CRs is clearly
seen. Initially, the CR protons are concentrated around the shocks,
and the proton spectrum is close to the spectrum of the isolated SNR
shock, $dN_p(p)/dp \propto 1/p^2$. As the inter-shock distance
reduces, the proton spectrum gets harder and the CRs concentrate  in
between the shocks. At the distance of 0.1 pc, the spectrum of CR
protons accelerated in the two-shocks system almost coincides with
the spectrum $dN(p)/dp\propto 1/p$.

The later result is consistent with the expectation that for the
efficient acceleration process in such a system, the acceleration
time for the highest energy particles, $\tau_{a}$, should be
comparable to the dynamical time $\tau_{\rm dyn}$. The condition is
satisfied if: (i) $u_{sn}L_{12}>3\cdot D_{sn}$ and
$u_{sn}L_{12}>3\cdot D_{sw}$, where $u_{sn}$ - velocity of the SNR
shock, $D_{sn}$ and $D_{sw}$ - the Bohm-type diffusion coefficients
for the supernova and the stellar wind shocks correspondingly
\citep[see][]{byk11}, (ii) the diffusion coefficient in the
interstellar medium between the shocks $D_{is}\gg D_{sn},D_{sw}$. We
did not model here the magnetic field amplification process, but
rather parameterized the Bohm diffusion coefficients relying on the
simulations of DSA magnetic field amplification  at a single shock
performed by \citet[][]{VladByk08}. Assuming that the amplified
magnetic field near the shock is $\sim 100$ $\mu G$ as it was
inferred from young SNR observations recently reviewed by
\citet[][]{vink12}, the distance between the shock fronts $L_{12}
\sim 1$ pc, we used $D_{is}=100\times D_{sn}$ and $D_{sw}\simeq
D_{sn}$ for the Bohm-type diffusion.

\section{Discussion}
The simplified non-linear model of particle acceleration in a
converging flow between a young supernova shell and a fast wind of a
massive early type star predicts a hard spectrum of CR protons
confined in the flow and CRs escaping the accelerator of about a
decade width in energy as shown in Fig.~\ref{fig:fig2}. In the
considered case of the SNR shock and wind velocities of $\sim
3,000~\kms$ and about a parsec distance between the shocks the
maximal energies accelerated protons can reach 10$^5$ GeV. The
energy is close to the expected DSA limit in an isolated supernova
shock \citep[see e.g.][]{lc83,Hillas05}. In individual SNRs the
maximal energy depends strongly on the circumstellar matter of the
SN progenitor and  the maximal energy of the accelerated CRs can be
reached typically at the early Sedov phase. It is important that if
SNR is expanding in a cluster of young massive stars both the
maximal energy  and the flux of the escaping accelerated CRs may be
non-monotonous functions of the SN age, and may have secondary
maxima at the moments of close approaches of the SNR and fast
stellar winds. It should be noted that massive shells that are
surrounding the stellar winds \citep[see e.g.][]{winds99} may be
swept away by the first few SN explosions in a compact cluster of
young massive stars, thus alleviating the collisions of supernova
shells with the wind termination shocks. The stage of the closely
approaching SNR and stellar wind flows that is favorable for
particle acceleration typically lasts for 300-1,000 years depending
on the stellar wind velocity and the termination shock radius.
Therefore, such sources can likely contribute to the galactic CRs
population. The colliding supersonic flows with two shocks also
occur in the early-type runaway stars. Particle acceleration in a
way similar to that discussed above may occur in between the
bow-shock and the wind termination shock of a fast runaway early
type star. Recently a bow-shock survey with 28 fast runaway OB stars
was compiled by \citet[][]{periea12} providing candidate source for
particle accelerators. The velocities of runaway early type stars
are well below that of the supernova shocks discussed above and
therefore the maximal energies of accelerated particles are expected
to be lower than that in SNR-SW collision. However the fast runaway
early type stars may still be sources of non-thermal emission with
hard photon spectra.

\begin{figure}
\resizebox{10cm}{10cm}{
\includegraphics{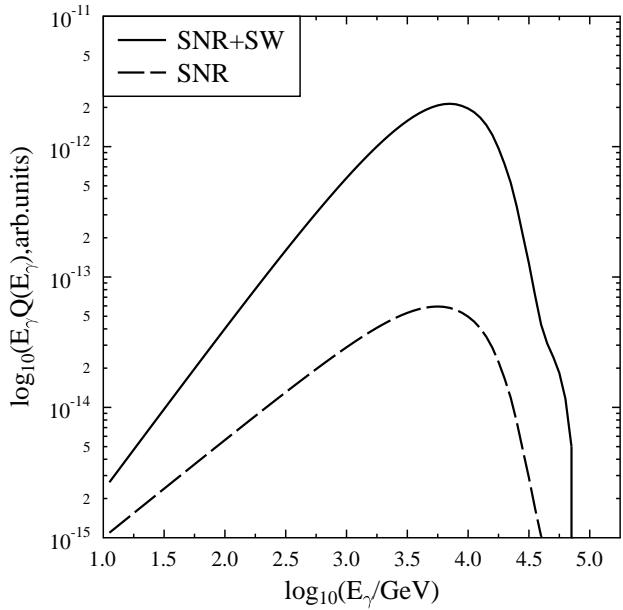}}
\caption{Model spectral energy distribution (SED) of the inverse
Compton emission from electrons accelerated at the colliding shock
flow (solid line) to compare with the emission of the electrons
accelerated by a single isolated shock of the same speed (dashed
line). The SEDs are given in arbitrary units and calculated for the
intershock distance of 0.1 pc. The corresponding distribution
function of the IC emitting electrons is shown in the bottom panel
in Fig.~\ref{fig:fig3}} \label{fig:fig4}
\end{figure}

The simplified non-linear model discussed in the paper predicts a
hard high energy end of CR particle spectrum. The simulated spectrum
of electrons is shown in Fig.~\ref{fig:fig3} (dashed curve). The
shape of the electron spectrum is similar to that of protons, but
the synchrotron/IC cooling effects result in the spectrum cut-off at
lower energies. The electromagnetic radiation produced by hard
spectra of electron/positron confined in the accelerator can be
potentially observed, thus constraining the physical parameters of
the observed objects. In Fig.~\ref{fig:fig4} we present a model
spectrum of the inverse Compton (IC) emission from the electrons
accelerated in a colliding shock flow comparing to the IC spectrum
of a single shock flow modeled earlier \citep[see
e.g.][]{Gaisser98,baringea99,bceu00}. To construct the IC spectrum
we used the same emission model which was applied to model high
energy emission of the supernova remnant IC~443 described in detail
in \citet{bceu00}. SNR IC~443 is likely located in Gem OB1 - young
massive star association. Therefore in addition to the local
interstellar photon spectrum which following \citet[][]{Mathis83}
was approximated by the sum of diluted blackbody spectra with
temperatures 2.7 K, 7500 K, 4000 K and 3000 K a major infrared
component resulting from the dust heated in OB association must be
accounted for. In the vicinity of IC~443 according to {\sl IRAS}
observations of \citet[][]{Saken92} we approximated the infrared
component with two temperatures fit (185 K and 34.3 K components).
If the shock collision region is located at the distance further
than about 2 parsecs from the young massive star then the UV
emission of the star does not dominate the IC losses of relativistic
electrons.

The electron distribution used in the IC spectrum computation is
shown in the bottom panel in Fig.~\ref{fig:fig3}. The
electron/positron distribution function $f_{e}(x,p,t)$  was computed
with the time-dependent model Eq.(\ref{difadv_e}) that accounts for
non-linear flow modification in the shocks vicinity as discussed in
the previous section, and the electron synchrotron/IC losses
included. It is clearly seen that due to the hard spectrum of
accelerated electrons up to $10^{4}$ GeV the emissivity of the
SNR-SW system is higher than that for the single SNR shock.
TeV-sources with very hard spectra are not always easily identified
with their GeV counterparts and some of these may comprise the
population of so-called "dark accelerators". The prototype of the
dark accelerators was TeV J2032+4130 discovered a decade ago by
\citet[][]{aharCygOB_02}, though the source is possibly associated
with the pulsar 2FGL J2032.2+4126 detected later in GeV regime by
Fermi observatory. Another high energy source was found by H.E.S.S.
\citep[see][]{hess_Wd1} in the vicinity of the young massive stellar
cluster Westerlund 1. The sources originating in the colliding
supersonic flows considered above may contribute to the VHE emission
of young massive star associations \citep[see, e.g.,][]{b01,tdr04}
    and the star-burst galaxies that are also known to be bright
gamma-ray sources \citep[see, e.g.,][]{hess_ngc253}. The main goal
of the paper is to discuss the unique features of particle
accelerators associated with the colliding shock flows. To model the
broadband non-thermal emission spectra of the sources one has to
discuss in detail matter and magnetic field distributions in the
complex flows. The structure of magnetic field in the winds of young
massive stars is under study, see for a review
\citet[][]{walderea12}. In the IC spectra modeling we assumed that
the CR electrons are the test particles that are propagating in the
flow modified by the CR protons. We did not consider here any
specific electron injection model to calculate their absolute fluxes
and therefore the IC spectra in Fig.~\ref{fig:fig4} are shown in
arbitrary units. We will present a more detailed emission spectra
modeling of colliding shock flows elsewhere.

\section*{Acknowledgments}
We thank the referee for constructive and useful comments. The work
was supported in part by the RAS Programs (P21 and OFN 16), and
also, by the Russian government grant 11.G34.31.0001 to the
Saint-Petersburg State Polytechnical University, by the RFBR grant
11-02-12082-ofi-m-2011 and by Ministry of Education and Science of
Russian Federation (Agreement No.8409, 2012). The numerical
simulations were performed at JSCC RAS and the SC at Ioffe
Institute. A.M.B thanks ISSI team 199 for discussion.

\bibliographystyle{mn2e}


\appendix

\section{Momentum-Energy flux conservation}\label{eneryC}
In this section we check the accuracy of the quasi-steady non-linear
model used, since it is relying on an the approximation of the CR
distribution function as given by Eq.~(\ref{eq:app}).  In a
steady-state system, mass, momentum and energy fluxes must be
constant in space obeying the following conservation laws
\begin{equation}
\rho(x)u(x)=\rho_{0}u_{0},
\end{equation}
\begin{equation}
\Phi_{P}(x)=\rho_{0}u_{0}^{2}+P_{g0},
\end{equation}
\begin{equation}
\Phi_{E}(x)=\Phi_{E0},
\end{equation}
where $\rho$ and $u$ are the mass density and the flow velocity,
$\Phi_{P}(x)$ and $\Phi_{E}(x)$ are the fluxes of the $x$-component
of the momentum  and energy in the $x$-direction. The  index "0"
marks the far upstream values.

\begin{figure}
\resizebox{10cm}{11cm}{
\includegraphics{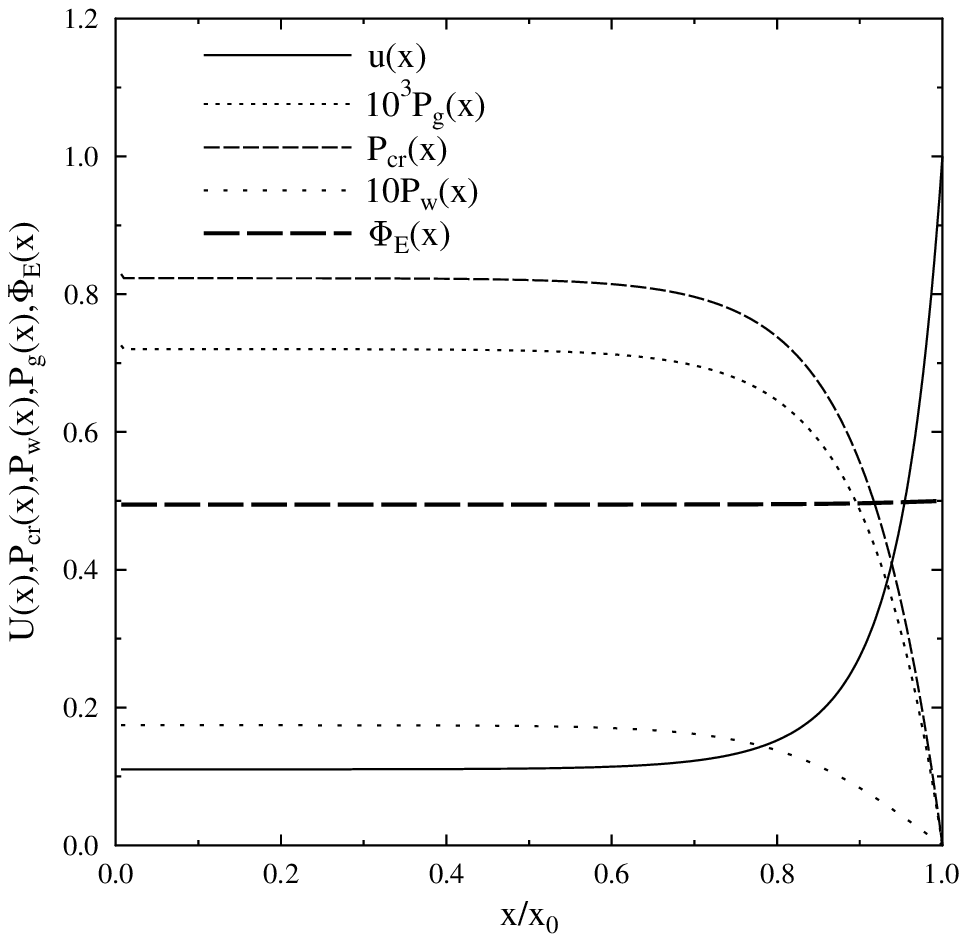}}
\caption{The energy conservation check. The shock velocity profile
$U(x) $ is given for the flow of Mach number $M_0=100$. The shock is
located at $x = 0$ and the position of the free escape boundary is
at $x = 1.0$. The cosmic ray pressure $P_{c}(x)$, the pressure of
the background gas $P_{g}(x)$ and the pressure of magnetic
fluctuations $P_{w}(x)$ are normalized to $\rho_0 u_{0}^2$. The
energy flux $\Phi_E(x)$ is normalized to $\rho_0 u_{0}^3$.}
\label{fig:energy}
\end{figure}

The CR energy and momentum fluxes are calculated from the CR
distribution function. In the paper we assumed that the CR energy
and momentum are dominated by protons, and the CR electrons are
treated as test particles that are propagating in the flow modified
by CR protons. A weakly anisotropic CR distribution function of the
accelerated particles can be approximated as
\begin{equation}
f(\textbf{r},p) = \frac{1}{4\pi}\left[ N (\textbf{r},p)+\frac{3}{v^{2}}\textbf{v}\textbf{J}(\textbf{r},p)\right],
\label{eq:f}
\end{equation}
where $N (\textbf{r},p)$ is the isotropic part of  the distribution
function, $\textbf{J}(\textbf{r},p) $ is the particles flux
\citep[e.g.][]{Topt85}. The $x$-component of the flux can be written
as:
\begin{equation}
J_{x}(x,p)= - D(x,p)\frac{\partial N}{\partial x} - \frac{p}{3}
\frac{\partial N}{\partial p} u(x)\, .
\end{equation}
The energy flux $\Phi_{cr}(x)$ is defined as
\begin{equation}
\Phi_{cr}(x)=\int K(p) v_{x}(p) f(x,p)d^{3}p\, ,
 \label{eq:phi}
\end{equation}
where $K(p)= E - mc^{2}$ is the kinetic energy of a particle,
$E=\sqrt{(pc)^{2}+m^{2}c^{4}}$ is the full energy of the particle,
$v_{x}=v \cdot cos\theta$, $\theta$  is the angle between the
particle momentum and the $x$-axis, and $v=\frac{\partial
E}{\partial p}$.

Substituting the $x$-component of the  distribution function
Eq.~(\ref{eq:f}) into the equation Eq.~(\ref{eq:phi}), one can
obtain
\begin{equation}
\Phi_{cr}(x)=\int\left[N (x,p)+\frac{3}{v} cos\theta
J_{x}(x,p)\right]\,K(p)\, v\, cos\theta\, \frac{d^{3}p}{4\pi} .
\end{equation}
Integrating over the angle one can get the CR
energy flux
\begin{equation}
\Phi_{cr}(x)=\int_{0}^{\infty}\,K(p) \left[- D(x,p)\frac{\partial
N}{\partial x} - \frac{p}{3} \frac{\partial N}{\partial p}
u\right]p^{2}dp .
\end{equation}
Then, finally, the expression for the CR kinetic energy flux is
\begin{eqnarray}
\Phi_{cr}(x)= \int_{0}^{\infty} p^{2}dp \left[- D(x,p)
\frac{\partial N}{\partial x}K(p) + (\frac{pv}{3}+K(p))Nu\right]  .
\label{eq:phix}
\end{eqnarray}
The energy flux of the magnetic turbulence is expressed by
\begin{equation}
\Phi_w(x)\simeq 3 u(x) p_w(x) ,
\end{equation}
where $p_w(x)=\rho_0 u^2_0P_w(x)$ is the pressure of the magnetic
turbulence generated via resonant streaming instability (see
Eq.~(\ref{eq:pw}) and \citet[][]{Cap09}).

Therefore, the energy conservation law for the steady non-linear
model can be written as
\begin{equation}
\Phi_{g}(x)+ \Phi_w(x)+\Phi_{cr}(x)- \Phi_{esc} = \Phi_{E}(x),
\end{equation}

\begin{equation}
\frac{1}{2}\rho(x) u^{3}(x) +3 u(x) p_w(x)+ \frac{\gamma}{\gamma-1}
u(x) p_{g}(x) + \Phi_{cr}(x) = const \label{eq:const}
\end{equation}
where $\Phi_{cr}(x)$ is calculated from Eq.~(\ref{eq:phix}),
$$
p_g(x)=\rho_0 u^2_0P_g(x)
$$ is the pressure of the background gas
and  $$
\Phi_{esc}=\int_{p_{inj}}^{\infty}E \cdot \phi_{esc}(p)dp
$$ is
the energy flux of the escaping CRs.

In  Figure~\ref{fig:energy} we illustrate the accuracy of the energy
conservation in the employed model of particle acceleration. The
dimensionless spatial coordinate is normalized to the FEB position
distance $x_0$. The energy flux conservation accuracy is about
$1\%$. Some small inconstancy of the full energy flux can be
attributed to the particular choice of the approximation of the CR
distribution function given by Eq.~(\ref{eq:app}) that was used in
the non-linear model.

\end{document}